\documentclass[prl,aps,twocolumn]{revtex4}
\usepackage{graphicx}
\usepackage{bm}
\usepackage{amsmath}
\begin{document}

\title{Intramolecular vibrational energy redistribution
as state space diffusion: Classical-Quantum correspondence}

\author{Aravindan Semparithi, and Srihari Keshavamurthy}

\affiliation{Department of Chemistry, Indian Institute
of Technology, Kanpur, Uttar Pradesh 208016, India}

\date{\today}

\begin{abstract}
We study the intramolecular vibrational energy redistribution (IVR)  
dynamics of an effective spectroscopic 
Hamiltonian describing the four coupled high frequency modes 
of CDBrClF. 
The IVR dynamics
ensuing from nearly isoenergetic zeroth-order states, an edge (overtone)
and an interior (combination) state, is
studied from a state space diffusion perspective. 
A wavelet based time-frequency analysis
reveals an inhomogeneous phase space
due to the trapping of classical trajectories.
Consequently the interior state has a smaller effective IVR dimension
as compared to the edge state.
\end{abstract}

\maketitle

Investigating the dynamics of an initially localized vibrational excitation
of a molecule in terms of timescales, final destinations and
competing pathways has been of considerable interest to chemical
physicists for a number of 
decades\cite{um91,Kel95,Ezr98,nf96,kp00,Gru00,gw04}. 
Due to the sustained theoretical\cite{um91,Kel95,Ezr98,gw04} 
and experimental efforts\cite{nf96,kp00,Gru00,gw04}
it is only now that a fairly detailed picture of the intramolecular
vibrational energy flow is beginning to emerge. 
Recent studies\cite{sw92,sw95,sww95,sww96} suggest that 
IVR can be described as a diffusion in
the zeroth-order quantum number space (also known as the state space) 
mediated by the anharmonic resonances coupling the zeroth-order states.
The state space approach\cite{Gru00} makes 
several predictions on the observables associated with IVR.
Foremost among them 
is that an initial zeroth-order bright state
$|b\rangle$ diffuses anisotropically on a 
manifold of dimension $D$ much smaller
than $N$ (or $N-1$ with energy constraint). $N$ is the number of vibrational
modes in the molecule.
As a result the survival probability
$P_{b}(t)$ exhibits power law behaviour on intermediate time scales
\begin{equation}
P_{b}(t) = |\langle b|b(t) \rangle|^{2} \sim
\sigma_{b} + 
(1-\sigma_{b})\left[1 + \frac{2t}{\tau D}\right]^{-D/2}
\end{equation}
with $\sigma_{b} = \sum_{\alpha} |\langle b|\alpha \rangle|^{4}$
being the dilution factor of the zeroth-order state $|b\rangle$
and $|\alpha \rangle$ denoting the eigenstates of the system.
Wong and Gruebele\cite{wg99} explained the power law behaviour
from the state space perspective by providing a perturbative estimate
for $D$ as:
\begin{equation}
D \approx D(n) = 
\frac{\Delta \ln \sum_{i} L_{ib}^{2}}{\Delta \ln n} 
\label{dimeq}
\end{equation}
with $n = |{\bf n}_{i}-{\bf n}_{b}|$ being the distance, in state space,
from the state $|b\rangle$ to other states $|i\rangle$ and
the sum is over all states $|i\rangle$ such that
$|{\bf n}_{i}-{\bf n}_{b}| \leq n$.
The zeroth-order quantum numbers $n_{k}$ are associated with the 
state $|k\rangle$ and the symbol $\Delta$ indicates a finite difference
evaluation of the dimension due to the discrete nature of the
state space. In practice
one chooses two different distances $n$ in the state space
and evaluates eq.~\ref{dimeq} and thus $D \approx D(n)$.
The quantity
\begin{equation}
N_{loc}(|b\rangle) \equiv \sum_{i} L_{ib}^{2} =
\sum_{i}\left[ 1 + 
\left( \frac{\Delta E^{0}_{ib}}{V_{ib}}\right)^{2} \right]^{-1}
\end{equation}
is a measure of the number of states locally coupled to $|b\rangle$.
The difference in the zeroth-order energies is denoted by
$ \Delta E^{0}_{ib} = E^{0}_{i} - E^{0}_{b}$
and $V_{ib} = \langle i|V|b\rangle$. 
Notice that in the strong coupling limit, $\Delta E^{0}_{ib} \ll V_{ib}$, 
$L_{ib} \approx  1$ whereas in the opposite limit
$L_{ib} \approx 0.$ Thus $D_{v}(n)$ can range between the full state space
dimension and zero\cite{wg99}. For further discussions on the origin and
approximations inherent to eq.~\ref{dimeq} we refer the reader to the
original reference\cite{wg99}. In the context of the present study it
is sufficient to note that $D \propto N_{loc}$ which has been confirmed
in the earlier work\cite{wg99}.

Clearly Eq.~\ref{dimeq}
explicitly includes the various anharmonic resonances 
and hence not only the local nature but the directionality
of the energy flow is also taken into account. 
The main point is that the above
estimate for $D$, which can be obtained without computing the actual
dynamics, is crucially dependent on the nature of the IVR diffusion in
the state space. However, to the best of our knowledge, precious little
is known about the dynamics associated with
the state space diffusion. 
Our motivation for investigating the
IVR dynamics in the state space has to do with the observation
that the state space model shares many of the important features found in
the classical-quantum correspondence
studies of IVR\cite{um91,Ezr98,Kel95}. 
Classical dynamical studies  
identify the nonlinear resonance network as the
crucial object. On such a network, directionality and the local nature of 
IVR arises rather naturally mainly due to the reason that molecular
phase spaces are mixed regular-chaotic even at fairly high energies. 
How does the mixed phase space influence the IVR
diffusion in the state space? Is there any relation, and hence correlation,
between the classical resonance network and the IVR dimension $D$ in the
state space? Is it possible that local dynamical traps in the classical
phase space can affect the validity of Eq.~\ref{dimeq}? 
Answers to these questions can have significant impact on our ability
to control IVR and hence reaction dynamics.
The issues involved are subtle and this preliminary work attempts to
address the questions by studying a specific system.

Although detailed
classical-quantum correspondence studies 
of IVR have been performed\cite{um91,Kel95,Ezr98,Kes99}
on systems with two degrees of freedom, in order to address
the questions one needs to analyze atleast a three degree of freedom
case. This is due to the fact that the scaling theory
of Schofield and Wolynes posits $D=2$ as the critical scaling {\it i.e.,}
near the IVR threshold\cite{sw92}. 
Thus for systems with two degrees of freedom the
separation of diffusive and critical regimes is not very sharp\cite{Kes99}.
However, studying IVR from the phase space perspective
is difficult in systems with three or more degrees of freedom. In this 
study we use a time-frequency technique 
proposed by Arevalo and Wiggins\cite{aw01}
to construct a useful phase space representation of the resonance
network for three degrees of freedom. 
Such an approach, as seen below and 
in many recent studies\cite{sk03,bhc05}, is
particularly well suited for our purpose.
Thus we choose an effective
spectroscopic Hamiltonian\cite{bhmqs00} describing the energy flow dynamics
between the four high frequency modes of CDBrClF. 
The Hamiltonian $H = H_{0} + V_{res}$ with
the anharmonic zeroth-order part
\begin{equation}
H_{0} = \sum_{j}\omega_{j} a_{j}^{\dagger} a_{j} +
\sum_{i \leq j} x_{ij} a_{i}^{\dagger} a_{i} a_{j}^{\dagger} a_{j}
\end{equation}
has various anharmonic resonances coupling
the four normal modes denoted by $s$ (CD-stretch), 
$f$ (CF-stretch) and
$(a,b)$ (CD-bending modes)
\begin{eqnarray}
V_{res} &=& \sum_{j \leq m}^{a,b,f}
\frac{k_{sjm}}{2\sqrt{2}} (a_{s}a_{j}^{\dagger}
a_{m}^{\dagger} + a_{s}^{\dagger} a_{j} a_{m}) \nonumber \\
&+& \frac{1}{2} \sum_{j < m}^{a,b,f}
\gamma_{jm} (a_{j}^{\dagger} a_{j}^{\dagger} a_{m} a_{m} +
a_{j} a_{j} a_{m}^{\dagger} a_{m}^{\dagger}) 
\label{resham}
\end{eqnarray}
The harmonic creation and destruction operators for the j$^{\rm th}$ mode
are denoted by $a_{j}^{\dagger}$ and $a_{j}$ respectively.
Note that despite having four coupled modes the system has effectively three
degrees of freedom due to the existence of a conserved polyad
$N = v_{s} + (v_{f} + v_{a} + v_{b})/2$.
In this work we choose $N=5$ for illustrating the main idea. Similar
results are seen in other systems and the details will be published
later. The values of the various parameters are taken from the fit
in reference\cite{bhmqs00} (fourth column, Table VIII).
The Fermi resonance strengths $k_{sjm}$ are
larger then the mean energy level spacings (13.7 cm$^{-1}$ $\sim$ 2.5 ps)
of $H_{0}$ for $N=5$.
Thus this
is an example of a strongly coupled system
and the multiple
Fermi resonances render the classical dynamics irregular.
We investigate the 
IVR dynamics out of two nearly isoenergetic 
zeroth-order states 
$|v_{s},v_{f},v_{a},v_{b} \rangle = |5000\rangle,$ and $|3301\rangle$
denoted for convenience as $|1\rangle$ and $|2\rangle$ respectively.
The experimentally accessible state $|1\rangle$
has energy $E^{0}_{1} \approx 10571$ cm$^{-1}$ whereas
the combination state has 
$E^{0}_{2} \approx 10567$ cm$^{-1} \approx E^{0}_{1}$.
We restrict our study to these two states although there are other close by
states within a mean level spacing.
In terms of their location in state space the state $|1\rangle$ is
an example of an edge state whereas
$|2\rangle$ is
an example of an interior state.

\begin{figure} [htbp]
\begin{center}
\includegraphics*[width=65mm]{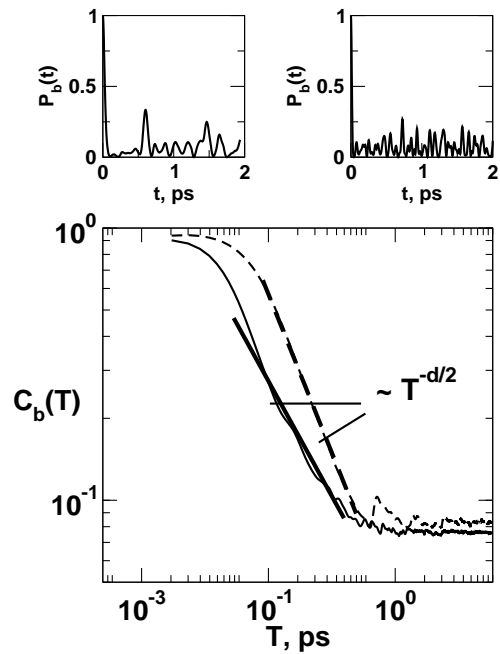}
\caption{Survival probabilities $P_{b}(t)$ and time-smoothed survival
probabilities $C_{b}(T)$ for the two isoenergetic states.
The left and right top panels show $P_{b}(t)$ for
$|1\rangle$ and $|2\rangle$ respectively.
Intermediate time power law behaviour is exhibited by $C_{b}(T)$
in both cases, dashed line (state $|1\rangle$)
and solid line (state $|2\rangle$). Note that the $C_{b}(T)$ data is shown
for $T=10$ ps indicating small oscillations about the respective dilution
factors $\sigma_{b}$.}
\label{fig 1}
\end{center}
\end{figure}

Given the Hamiltonian and the resonant couplings the number of states 
coupled locally to $|1\rangle$ and $|2\rangle$ can be estimated as 
$N_{loc} = 1.8$, and $3.0$ respectively. Combined with the fact that
$N_{eff} \equiv \sigma_{b}^{-1} \approx 12$ one expects fast IVR
from both the states. Since $N_{loc}(|1\rangle) < N_{loc}(|2\rangle)$
the decay is much faster at short times for $|2\rangle$.
This is confirmed in
Fig.~\ref{fig 1} which shows $P_{b}(t)$ for the states.
However Fig.~\ref{fig 1} also shows the time-smoothed survival
probability\cite{kpg92,hs94}
\begin{equation}
C_{b}(T)= \frac{1}{T} \int_{0}^{T} dt P_{b}(t)
\label{tacor}
\end{equation}
associated with the states
and, importantly, highlights 
a power law behavior of $C_{b}(T) \sim T^{-d/2}$
at intermediate times - a sign of incomplete IVR\cite{wg99,gw04}.
Note that the persistent recurrences in $P_{b}(t)$ occur for much longer
times as evident from the results for $C_{b}(T)$.
Earlier works\cite{kpg92,hs94}, in an apparently different context,
have associated the power law behaviour with the multifractality of
the eigenstates and the local density of states.
The power law exponent or effective
IVR dimensionality in the state space are determined to be
$d_{1} \sim 1.8$ and $d_{2} \sim 1.3$ which are smaller than the
three dimensional state space. 
Incidentally, a purely exponential decay of $P_{b}(t)$ would imply
$C_{b}(T) \sim T^{-1}$ irrespective of the dimensionality of the
state space.
More surprising observation from Fig.~\ref{fig 1} is that the 
interior state shows faster short time IVR but at longer times,
despite $N^{(2)}_{loc} > N^{(1)}_{loc}$,
explores an IVR manifold of smaller dimension as compared to the edge state.
Infact based on $N_{loc}$ and the strong couplings one would infer the
opposite from Eq.~\ref{dimeq}.

\begin{figure} [htbp]
\begin{center}
\includegraphics*[width=65mm]{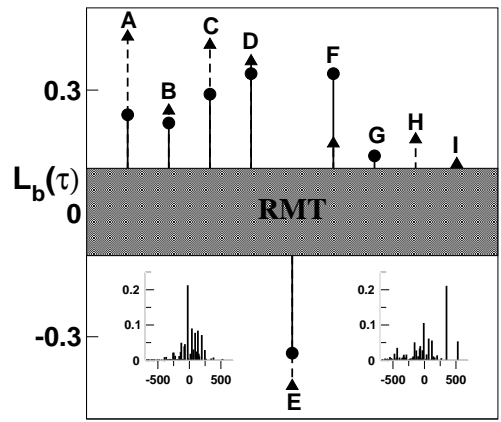}
\caption{Intensity-level velocity correlator $L_{b}(\tau)$ for
$|b\rangle = |1\rangle$ (circles) and $|2\rangle$ (triangles).
The random matrix (RMT) estimate is shown by the shaded region.
The parameter $\tau$ are taken to
be the various resonant strengths and the $sff,saa,sbb,sfa,sfb,sab,aabb,
ffaa,$ and $ffbb$ resonances are denoted by A,B,...,I respectively. The
averaging is performed over a range $\Delta E = \pm 700 cm^{-1}$ corresponding
to the width of the IVR feature as seen in the insets (Left inset corresponds
to $|1\rangle$ and the right inset to $|2\rangle$).}
\label{fig 2}
\end{center}
\end{figure}

Since $E^{0}_{1} \approx E^{0}_{2}$ the results in Fig.~\ref{fig 1}
suggest different IVR mechanisms for the two states. This can be established
by correlating the intensities $p_{b \alpha} = |\langle b|\alpha \rangle|^{2}$
with the parametric variation of eigenvalues $E_{\alpha}(\tau)$ {\it i.e,} the
intensity-level velocity
correlation function\cite{Tom96,kct02}: 
\begin{equation}
L_{b}(\tau) = \frac{1}{\sigma_{p}\sigma_{v}}\left\langle p_{b \alpha}
\frac{\partial E_{\alpha}}{\partial \tau} \right \rangle_{\Delta E}
\end{equation}
where $\sigma_{p}$ and $\sigma_{v}$ denote the intensity and
level velocity variances respectively. The parameter $\tau$ corresponds to
the resonant coupling strengths in eq.~\ref{resham}
and $\Delta E$ is the width of the IVR
feature.
Recent work\cite{kct02} has shown that
$L_{b}(\tau)$ can identify the dominant resonances
that control the IVR dynamics.
In Fig.~\ref{fig 2} we show the correlator
$L_{b}(\tau)$ for $|1\rangle$
and $|2\rangle$. 
Random matrix theory (RMT) predicts\cite{Tom96}
$L_{b}(\tau) \sim 0 \pm 1/\sqrt{N}$ with $N$ being the number of
eigenstates under the IVR feature and hence
ergodicity implies 
a vanishing correlator for any state of choice.
It is clear from Fig.~\ref{fig 2} that several of the correlators
violate the RMT estimate indicating localization.
In particular $L_{b}(\tau)$ indicates differing IVR dynamics out
of $|1\rangle$ and $|2\rangle$. For instance, $L_{b}(k_{sff})$ for
the states differ by about $0.2$ which is greater than the fluctuations
allowed by RMT ($\approx 0.11$). 
Note that the results in Fig.~\ref{fig 2} 
support the local RMT approach\cite{lw90,lw97}, developed
by Logan, Leitner, and Wolynes, which is consistent with the power law
decay of $P_{b}(t)$ and thus $d \sim D$ (cf. eq.~\ref{dimeq}).

\begin{figure} [htbp]
\begin{center}
\includegraphics*[width=65mm]{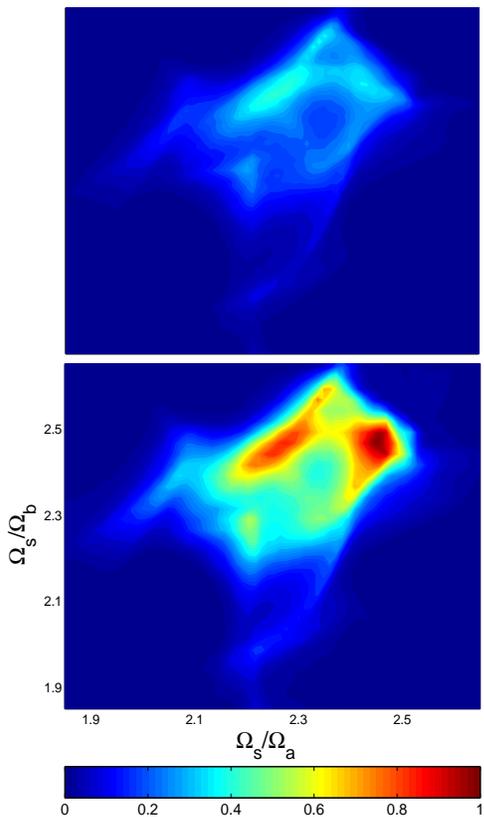}
\caption{Dynamical Arnol'd web plotted in the frequency ratio space
$(\Omega_{s}/\Omega_{a},\Omega_{s}/\Omega_{b})$
for $|1\rangle$ (top) and $|2\rangle$ (bottom).
The scale for the axes is identical for both the figures.
The data is
obtained by propagating $5000$ trajectories with fixed
initial actions corresponding to the state of interest and
varying angles such that $H({\bf I},{\bm \theta}) \approx
E^{0}_{b}$. The color scale (normalized) indicates the number of times
that a region is visited by the trajectories.
Total propagation time $T = 10$ ps and note that
the figures look similar already by about $T=2.5$ ps.}
\label{fig 3}
\end{center}
\end{figure}

We now show that the observed power law in Fig.~\ref{fig 1} and the
slower IVR dynamics of the interior state $|2\rangle$ are due to
the existence of dynamical traps in the classical phase space.
First the classical limit Hamiltonian $H({\bf I},{\bm \theta})$
is constructed using the correspondence\cite{Kel95} 
$a_{j} \rightarrow \sqrt{I_{j}} e^{i \theta_{j}}$ with
$({\bf I}, {\bm \theta})$ being the action-angle variables of $H_{0}$.
Next, classical trajectories with initial conditions such that
$H({\bf I},{\bm \theta}) \approx E_{b}^{0} $ and actions ${\bf I}$
restricted to the specific state are generated. 
For every trajectory the dynamical function
$z_{k}(t) = \sqrt{2I_{k}(t)} \exp(i\theta_{k}(t))$ with
$k=s,f,a,b$ is subjected to the
wavelet transform\cite{aw01}:
\begin{equation}
W_{g}z_{k}(A,B) = A^{-1/2} \int_{-\infty}^{\infty} z_{k}(t)
g^{*}\left(\frac{t-B}{A}\right) dt
\label{wave}
\end{equation}
with $A > 0$ and real $B$. The function 
$g(t) = (2\pi \sigma^{2})^{-1/2} \exp(2\pi i \lambda t - t^{2}/\sigma^{2})$ 
is taken to be the simple
Morlet-Grossman wavelet\cite{aw01} with $\lambda=1$ and $\sigma=2$.
Eq.~\ref{wave} yields the frequency
content of $z_{k}(t)$ over a time window around $t=B$.
In this work we obtain the local frequency associated with $z_{k}(t)$
by determining the scale ($A$, inversely proportional to frequency) 
which maximizes the modulus of the
wavelet transform {\it i.e.,} 
$\Omega_{k}(t=B) = {\rm max}_{A}|W_{g}z_{k}(A,B)|$.
This gives the nonlinear frequencies ${\bm \Omega}(t)$ and the
dynamics at $E = E^{0}_{b}$ is followed in 
the frequency ratio space\cite{mde87} 
$(\Omega_{s}/\Omega_{a},\Omega_{s}/\Omega_{b})$.
The frequency ratio space is divided into cells and the
number of times that a cell is visited for all the trajectories
gives the density plot. We further normalize the highest
density to one for convenience. 
Two points should be noted at this stage. 
First, such a density plot is providing a glimpse
of the energy shell and is reflecting the
full dynamics including the important resonances.
Thus we are mapping out parts of the Arnol'd web {\it i.e.,} the resonance
network that is actually utilized by the system. 
Secondly, we are computing
a slice of the energy shell and
for strongly coupled systems
one expects the phase space structure to be 
different for different slices {\it i.e.,} nontrivial dependence on
the angles ${\bm \theta}$.
We thus compute a highly averaged structure in the frequency ratio space
which is nevertheless still capable of providing important information
on the nature of the classical dynamics.
The resulting density plots are
shown in Fig.~\ref{fig 3}
for $|1\rangle $ and $|2\rangle$ and look similar because $E^{0}_{1} \approx
E^{0}_{2}$. 

Fig.~\ref{fig 3} clearly shows the heterogeneous
or nonuniform nature of the density despite angle averaging. 
This suggests that at $E \approx E_{b}^{0}$
there are dynamical traps in the phase space and hence
the dynamics is nonergodic. 
However more important is the nature of these
trapping regions since one expects them to provide insights into
the IVR dynamics.
In Fig.~\ref{fig 3} two significant traps
corresponding to $\Omega_{s} \approx 2 \Omega_{f}$ ($sff$) 
and another to $\Omega_{a} \approx \Omega_{b}$ ($ab$) are observed.
Note that the $\Omega_{a} \approx \Omega_{b}$ lock is an induced effect
and in particular persists upon removing the $\gamma_{ab}$ term from
eq.~\ref{resham}. The traps are seen for both states, hence the
power law behavior of $C_{b}(T)$ for both states, 
but the extent of trappings is different.
The $ab$-lock is more extensive
for the state $|2\rangle$ as opposed to the state $|1\rangle$.
Given the extensive $ab$-lock for the dynamics associated with
$|2\rangle$ one imagines that the
CD-bend modes get isolated rather quickly
from the other two modes. In other words, as soon as the energy flows into
one of the bends the other bend starts to resonantly shuttle this energy
back and forth resulting in restricted IVR. This correlates well
with the results in Fig.~\ref{fig 1} which shows a smaller effective
dimensionality of the IVR manifold for $|2\rangle$. 
Thus one can infer that
the restricted IVR for the interior state is due to the extensive
trapping in the classical phase space. 
The effective dimension of the IVR manifold $d$ arising due to a
power law behavior of the quantum $C_{b}(T)$ indicated restricted IVR.
At the same time analysis of the classical dynamics
shows the heterogeneous nature of the phase space due to resonance
trappings. If the density plots look homogeneous due to the absence of
any trapping regions then one can associate a
dimensionality $d_{fr} = 2$ to
the frequency space. However Fig.~\ref{fig 3} show that
for both states
$d_{fr} < 2$ and one might therefore associate a fractal
dimension between $d_{fr} = 1$ and $d_{fr} = 2$. 
Clearly $d_{fr}(1) > d_{fr}(2)$ and hence
one can conjecture that $d \sim d_{fr}$ {\it i.e.,} the effective
dimensionality of the IVR manifold is the same as the effective
dimensionality of the frequency ratio space or resonance web. 

We conclude by making a few observations. 
Gambogi {\it et al.} observed\cite{gtls93} a similar effect in propyne
wherein the eigenstate-resolved spectra indicated that the combination
mode $v_{1}+2v_{6}$ is much less perturbed by IVR as compared to the
nearly isoenergetic $3v_{1}$ overtone state.
It was argued that such effects are to be expected in large molecules.
The present example shows that enhanced instability of overtone states
as compared to the combination states can occur in few mode systems as well.
The current study highlights this to be a dynamical effect.
The decoupling
of the $(s,f)$ modes from the $(a,b)$ modes for state $|2\rangle$
implies that the full Hamiltonian for $E \approx E^{0}_{b}$ is
dynamically decoupled into two sub-Hamiltonians: one approximately
conserving the polyad $2v_{s}+v_{f}$ and the other conserving the
polyad $v_{a}+v_{b}$. The precise forms of such sub-Hamiltonians is
not clear as of now.
Finally, the extensive $ab$-lock and the resulting decoupling of
the CD-bend modes may relate to the observation made by Beil {\it et al.}
on a possible case of an approximate symmetry\cite{bhmqs00} which
arises from a near conservation of a formal $a'$ symmetry associated
with the bending states. This point, however, requires further studies.

\end{document}